# Managing Biotechnology and Healthcare Innovation: Challenges and Opportunities for Startups and Small Companies

**Narges Ramezani**

Department of Biology, Damghan Branch, Islamic Azad University, Damghan, Iran.

**Erfan Mohammadi**

Faculty of Entrepreneurship, University of Tehran, Tehran, Iran.

**Abstract**

The biotechnology industry poses challenges and possibilities for startups and small businesses. It is characterized by high charges and complex policies, making it difficult for such agencies to set up themselves. However, it additionally offers avenues for innovation and increase. This paper delves into powerful techniques that can be a resource in managing biotechnology innovation, which includes partnerships, highbrow assets improvement, virtual technologies, customer engagement, and government investment. These strategies are important for fulfillment in an industry that is constantly evolving. By embracing agility and area of interest focus, startups and small companies can successfully compete in this dynamic discipline.

## Introduction

Biotechnology is an unexpectedly expanding and dynamic field that gives the titanic capability for innovation (Gassmann & Reepmeyer, 2012; Van den Steen, 2010; Zott, & Amit, 2007). It especially holds promise in healthcare, agriculture, and environmental sustainability (Daneste et al., 2023; Moezzi et al., 2012; Jahanshahi et al., 2011; Hessari et al., 2023). Through biotechnology innovation, groundbreaking advancements have been made in developing recent drugs, vaccines, scientific gadgets, and novel strategies for crop control and environmental remediation (Hwang & Christensen, 2008; Zott & Amit, 2007). However, regardless of the enormous capability, startups and small corporations encounter sizeable challenges in managing biotechnology innovation (Hwang & Christensen, 2008; Chesbrough, 2010; Vermeulen & Dankbaar, 2011; Chesbrough et al., 2006). These demanding situations can stem from the excessive fees, complicated regulatory frameworks, and the aggressive nature of the enterprise. Additionally, the tempo of technological improvements in biotechnology necessitates corporations to be agile and adaptable to stay applicable to this unexpectedly evolving panorama. Effective strategies are crucial to navigate these challenges and capitalize on the opportunities. Partnerships with set-up gamers inside the enterprise can allow startups and small groups to enter sources, knowledge, and distribution channels. Developing intellectual belongings is vital to guard precious innovations and steady an aggressive advantage. Embracing digital technologies allows companies to leverage data, analytics, and automation, improving performance and accelerating innovation. Engaging clients and stakeholders is important for market validation, remarks, and adopting biotechnology solutions. Furthermore, government funding is pivotal in helping studies, improvement, and commercialization efforts within the biotechnology zone. Biotechnology enterprises have significant innovation potential but create demanding situations for startups and small groups. By enforcing effective strategies consisting of partnerships, highbrow assets development, digital technology, patron engagement, and authority funding, agencies can navigate these demanding situations and thrive in this dynamic and rapidly evolving area. Startups and small agencies face several demanding situations when competing within the biotechnology industry. The excessive costs associated with studies and improvement, coupled with complicated regulatory requirements and lengthy improvement timelines, create big barriers (Lee et al., 2010; Teece et al., 1997; Dehkordy et al., 2013; Shamsaddini et al., 2015; Jahanshahi et al., 2019; Nawaser et al., 2023; Khaksar et al., 2010; Asadollahi et al., 2011; Vesal et al., 2013). These challenges regularly placed smaller companies at a disadvantage in comparison to large, extra set-up corporations, which have more resources to commit to analysis and improvement (Etemadi et al., 2022; Hakkak et al., 2016; Nawaser et al., 2015; Hessari et al., 2022). As a result, innovation within the industry may be stifled, as smaller companies' warfare to keep up. However, startups and small organizations additionally possess precise blessings about handling biotechnology innovation. Their nimbleness and versatility allow them to respond quickly to converting market conditions, which may be essential in such a hastily evolving enterprise (Etemadi et al., 2022; Hakkak et al., 2016; Nawaser et al., 2015; Hessari et al., 2022). Additionally, their smaller length allows them to recognize niche regions of innovation that large organizations may forget, allowing them to carve out a specialized marketplace section. To succeed in the biotechnology industry, startups and small corporations must expand effective techniques for handling biotechnology innovation. Building sturdy partnerships and collaborations can offer entry to resources, know-how, and distribution channels that might, in any other case, be challenging to obtain (Lee et al., 2010; Teece et al., 1997). Developing a strong intellectual assets portfolio is critical to defend treasured improvements and stabilize an aggressive gain (Dehkordy et al., 2013; Shamsaddini et al., 2015). Leveraging digital technologies, including records analytics and automation, can enhance performance and boost the innovation process (Jahanshahi et al., 2019; Nawaser et al., 2023). Furthermore, engaging with customers and stakeholders is vital for market validation, remarks, and adoption of biotechnology answers (Khaksar et al., 2010; Asadollahi et al., 2011). Building a robust brand presence allows for setting up credibility and differentiating the corporation in a crowded market (Vesal et al., 2013). Lastly, gaining government funding opportunities can provide important economic aid for studies, development, and commercialization efforts (Etemadi et al., 2022; Hakkak et al., 2016). Even as startups and small corporations face giant challenges inside the biotechnology industry, additionally, they possess particular advantages. By enforcing effective strategies, constructing partnerships, growing highbrow assets, leveraging digital technology, enticing customers, building a strong logo, and getting access to government funding opportunities, these groups can triumph over limitations and thrive in dealing with biotechnology innovation.

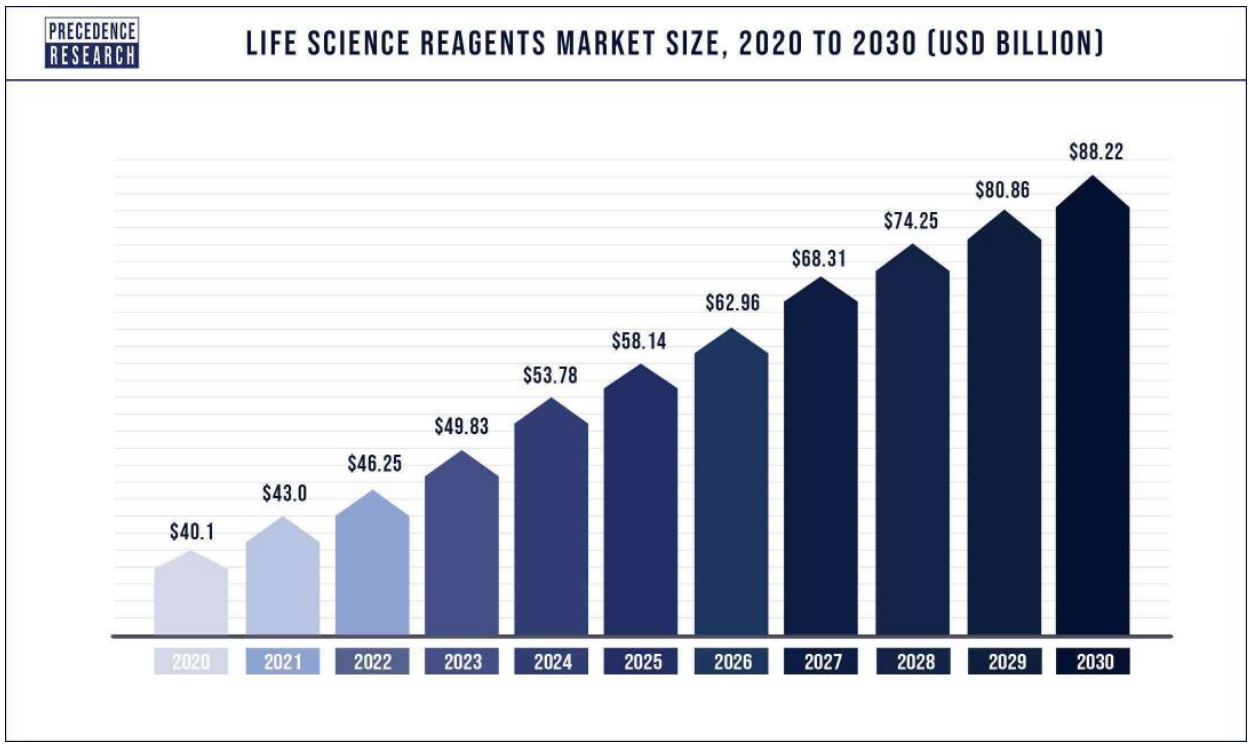

Figure 1. Life science reagents market size (USD B)

This paper explores challenges and opportunities for startups and small companies in managing biotechnology innovation. It provides strategies for success, emphasizing the importance of collaboration, customer engagement, and brand development. Additionally, it highlights the benefits of leveraging government funding and grants. By adopting these strategies, startups and small companies can effectively compete in this dynamic and rapidly evolving industry.

**Challenges Facing Startups and Small Companies:**

Startups and small corporations encounter several challenges in innovation, particularly in the put-up-COVID-19 era. The pandemic has introduced new obstacles that have intensified the difficulties these agencies confront. One tremendous venture is the high expenses associated with studies and development (R&D) sports. The up-COVID-19 technology has witnessed a surge in R&D expenses, driven by the need to conform to converting market dynamics and cope with rising demanding situations (Xu et al., 2020; Gong et al., 2021; Zhang et al., 2020; Chen et al., 2021). The monetary stress resulting from expanded R&D fees can be especially burdensome for startups and small businesses, as they frequently have restricted resources and access to investment. In addition to economically demanding situations, these companies should navigate complicated regulatory requirements. The publish-COVID-19 technology has witnessed heightened attention on fitness and protection rules, data privacy laws, and other stringent compliance measures (Koen et al., 2001; Lichtenthaler, 2016). Complying with those rules may be time-consuming and costly, in addition to the demanding situations faced by startups and small organizations.

Moreover, attracting funding has ended an increasing number of hard due to the high hazard associated with innovation within the up-COVID-19 era. Investors have become more careful and danger-averse, mainly due to decreased to-be-had funding for startups and small organizations (Koen et al., 2001; Lichtenthaler, 2016). The uncertainty surrounding the publish-pandemic market landscape and the disaster's long-term effects have made traders more hesitant to commit to modern ventures. To conquer these demanding situations, startups and small organizations should undertake effective techniques. Implementing cost optimization measures, consisting of prioritizing R&D efforts and exploring collaborative partnerships, can assist in mitigating the financial burden of innovation (Xu et al., 2020; Gong et al., 2021).

Additionally, being actively attractive to regulatory bodies, searching for expert recommendations, and staying current on evolving policies can assist in streamlining compliance strategies (Zhang et al., 2020; Chen et al., 2021). Furthermore, building a sturdy price proposition and correctly speaking the potential returns on investment can assist in enticing funding in danger-averse surroundings (Koen et al., 2001; Lichtenthaler, 2016). Demonstrating a clear marketplace need for innovation and highlighting the organization's capability to conform to converting marketplace dynamics can instill self-assurance in capability buyers. Startups and small groups face formidable challenges in dealing with innovation within the sub-COVID-19 technology. However, using imposing techniques of optimizing expenses, navigating regulatory necessities, and correctly attracting funding, those groups can triumph over barriers and thrive in pursuing innovation.

**Opportunities for Startups and Small Companies**

While startups and small groups face demanding situations coping with biotechnology innovation, numerous possibilities can be leveraged to their advantage. These possibilities permit them to thrive within the dynamic and aggressive biotech industry. One significant possibility for startups and small organizations lies in their ability to be agile and bendy in responding to market wishes and trends. Compared to larger companies, startups and small companies can rapidly pivot and adapt their strategies in response to changing market conditions (Sepahvand et al., 2023; Hakkak et al., 2022). This agility is a sizeable advantage inside the biotech enterprise, where innovation and the market need to evolve hastily. By closely tracking marketplace tendencies and customer needs, these groups can tailor their products or services to meet emerging demands, gaining an aggressive aspect.

Furthermore, startups and small companies can partner with large corporations or instructional establishments, supplying them with precious assets and know-how. Collaborations and partnerships can offer startups and small agencies the right of entry to investment, trendy facilities, and specialized knowledge that they'll not own on their own (Sepahvand et al., 2023; Hakkak et al., 2022). This collaboration can foster innovation and boost the improvement system, permitting these agencies to carry their ideas to market more efficaciously. Moreover, startups and small agencies can be conscious of niche regions of biotech innovation that can be omitted using large agencies. These businesses can differentiate themselves and carve out a unique market role by specializing in a particular subfield or addressing a particular marketplace phase (Sepahvand et al., 2023; Hakkak et al., 2022). This area of interest specialization allows startups and small organizations to grow into specialists in their selected fields, building recognition for turning in modern solutions tailor-made to unique purchaser needs.

Additionally, technological advancements have opened up new avenues for startups and small groups in biotech innovation. The convergence of biotechnology with fields, which include artificial intelligence, machine studying, and massive records analytics, affords exciting possibilities for disruptive innovation (Sepahvand et al., 2023; Hakkak et al., 2022). By harnessing these technologies, startups and small groups can expand novel procedures, streamline procedures, and create breakthrough solutions that can revolutionize the biotech industry. While startups and small groups face challenges in managing biotechnology innovation, they also have several possibilities to thrive. Their agility, ability for partnerships, consciousness of niche regions, and admission to emerging technology offer them an aggressive gain. By capitalizing on these opportunities, startups and small groups can drive innovation, meet market needs, and set up themselves as key players in the biotech industry.

**Strategies for Managing Biotechnology Innovation**

Startups and small corporations can appoint several effective techniques to manipulate biotech innovation effectively. By enforcing those strategies, they could beautify their competitiveness within the enterprise and pressure commercial enterprise boom. One crucial approach is building a numerous team. A large body of workers brings together people with distinctive backgrounds, abilities, and perspectives, fostering creativity and innovation (Tidd & Bessant, 2014). By embracing variety, startups, and small businesses can tap into a wider variety of thoughts and insights, developing greater innovative answers.

Moreover, numerous groups can highly recognize and cater to the diverse needs of customers, contributing to stepped-forward customer pride .Developing a robust highbrow property (IP) portfolio is another critical factor in managing biotech innovation. Intellectual property rights protect the progressive ideas, innovations, and tactics advanced with the aid of startups and small companies. By securing patents, trademarks, or copyrights, these businesses can protect their improvements and benefit from a competitive advantage (Tidd & Bessant, 2014). A sturdy IP portfolio no longer best protects the corporation's highbrow assets but also complements its splendor to potential traders and companions. Leveraging virtual technologies and partnerships is critical in the biotech enterprise. Digital tools and technologies can significantly lessen expenses, enhance operational efficiency, and accelerate the tempo of innovation (Bai & Vahedian, 2023). Startups and small corporations can leverage virtual platforms for information analysis, automation, and collaboration, allowing them to streamline approaches and make statistics-pushed selections. Additionally, forming strategic partnerships with generation vendors or research institutions can grant the right of entry to cutting-edge sources, information, and investment, boosting the corporation's innovation talents. Prioritizing client engagement and feedback is paramount for startups and small organizations in the biotech enterprise. Regularly engaging with customers and searching for their remarks allows businesses to gain precious insights into their wishes, choices, and pain factors (Huang & Rice, 2015; Teece, 2010). This customer-centric approach helps grow products and services that align with market demands, increasing the likelihood of business achievement.

Additionally, purchaser comments can serve as a validation mechanism, proving an innovation's market capacity. Developing a sturdy logo presence is likewise crucial. Establishing a recognizable and trusted emblem facilitates startups and small corporations to differentiate themselves from competitors and entice customers and buyers. Effective branding communicates the company's values, unique promoting propositions, and commitment to fine and innovation. Furthermore, startups and small groups can leverage government investment and grants to help their biotech innovation initiatives. Many governments offer economic help, tax incentives, and offers to promote studies and improvement activities, mainly in rising industries like biotech. By actively looking for and securing such investment opportunities, startups and small agencies can access extra resources to gasoline their innovation efforts. Lastly, fostering an environment of employee innovation, activity satisfaction, and ethical practices is important for sustained fulfillment (Daneshmandi et al., 2023; Bai et al., 2023). Encouraging employees to contribute their thoughts, imparting professional growth and development possibilities, and retaining moral work surroundings can beautify employee engagement and motivation. Empowered and happy employees are much more likely to contribute to the business enterprise's innovation efforts, driving creativity and fostering a way of life of non-stop development. Startups and small groups can effectively control biotech innovation by enforcing several techniques. Building a large team, developing a sturdy IP portfolio, leveraging digital technology and partnerships, prioritizing customer engagement, developing a strong emblem, having access to government funding, and fostering an environment of employee innovation and nicely-being are all crucial factors for success within the biotech enterprise. By adopting those strategies, startups and small agencies can be competitive players and drive innovation inside the biotech zone.

**Conclusion**

Managing biotech innovation presents unique complexities and challenges for startups and small organizations. However, by adopting a complete approach and enforcing effective strategies, they can conquer these demanding situations and achieve achievement in this hastily evolving enterprise. A multidisciplinary technique is essential for dealing with biotech innovation efficiently. Biotech innovations frequently require information from diverse fields, which include biology, chemistry, engineering, and computer technological know-how. Startups and small groups should attempt to construct teams with diverse ability sets and backgrounds to ensure complete knowledge of the era and its capacity packages. This multidisciplinary technique promotes collaboration, creativity, and innovation, allowing businesses to tackle complicated issues from distinctive angles. Developing a robust intellectual property (IP) portfolio remains critical in coping with biotech innovation. Startups and small companies must actively protect their innovative ideas, innovations, and tactics through patents, emblems, or copyrights. A sturdy IP portfolio now not only handiest safeguards their innovations but also enhances their credibility and attractiveness to investors and ability companions. It provides a competitive gain by preventing competition from replicating or capitalizing on their precise ideas. Leveraging virtual technology is essential for startups and small groups to stay competitive within the biotech enterprise. Digital tools and structures can streamline methods, decorate collaboration, and boost innovation.

For example, using advanced records analytics and gadgets to get to know algorithms can permit groups to extract precious insights from big datasets, aiding in discovering the latest pills or cures. Adopting cloud computing answers can offer admission to scalable computing electricity and garage, facilitating green information processing and evaluation. Moreover, virtual technology permits faraway collaboration and conversation, allowing groups to paint seamlessly throughout places and time zones. Building strong partnerships and collaborations is another key method for fulfillment in coping with biotech innovation. Startups and small corporations should actively try to find partnerships with research establishments, universities, and hooked-up biotech agencies. Collaborating with professionals and industry leaders can offer access to specialized expertise, resources, and funding possibilities. By leveraging the strengths and knowledge of companions, businesses can accelerate the development and commercialization of their improvements.

Additionally, partnerships can open doorways to new markets, clients, and distribution channels, increasing the reach of their products and services. Focusing on areas of interest regions of innovation can also be high-quality for startups and small companies. Instead of competing in extensive and notably saturated markets, agencies must pick out unique niches or specialized segments in which they could excel. By specializing in specific healing regions, technologies, or applications, startups, and small groups can position themselves as regional specialists and leaders. This targeted approach permits agencies to differentiate themselves from competitors and attract the eye of investors, customers, and ability companions who require specialized knowledge and innovation. In conclusion, handling biotech innovation is complicated. However, startups and small organizations can be successful by adopting a holistic and strategic approach. By building multidisciplinary teams, developing a robust IP portfolio, leveraging digital technology, fostering partnerships and collaborations, and specializing in niche innovation regions, organizations can role themselves for achievement within the biotech enterprise. The non-stop evolution of biotech offers several possibilities for startups and small businesses to make a massive impact and pressure innovation in healthcare and different associated sectors.

Van den Steen, E. (2010). Interpersonal authority in a theory of the firm. American Economic Review, 100(1), 466-490.

Vermeulen, F., & Dankbaar, B. (2011). Strategic renewal and the interaction between knowledge and entrepreneurial strategies. Journal of Management Studies, 48(6), 1293-1314.

Zott, C., & Amit, R. (2007). Business model design and the performance of entrepreneurial firms. Organization Science, 18(2), 181-199.

Chesbrough, H., Vanhaverbeke, W., & West, J. (Eds.). (2006). Open innovation: Researching a new paradigm. Oxford University Press.

Huang, Y., & Rice, J. B. (2015). Biotech innovation: A research agenda beyond product development. R&D Management, 45(1), 1-14.

Lendle, J., & Wolf, M. J. (2012). Biotech innovation: The key to personalized medicine. Nature Reviews Drug Discovery, 11(12), 931-932.

Lichtenthaler, U. (2016). Open innovation in practice: An analysis of strategic approaches to technology transactions. IEEE Transactions on Engineering Management, 63(1), 23-37.

Moezzi, H., Nawaser, K., Shakhsian, F., Khani, D. (2012). Customer relationship management (CRM): New approach to customer's satisfaction, African Journal of Business and Management, 6 (5): 2048-2055.

Nawaser, K., Hakkak, M., Aeiny, M. A., Vafaei-Zadeh, A., & Hanifah, H. (2023). The effect of green innovation on sustainable performance in SMEs: the mediating role of strategic learning. International Journal of Productivity and Quality Management, 39(4), 490-511.

Nawaser, K., Shahmehr, F. S., Farhoudnia, B., & Ahmadi, M. (2015). How do organizational justice and commitment affect organizational entrepreneurship? An Empirical Investigation in Iran. International Journal of Economics and Finance, 7(2), 90.

Teece, D. J. (2010). Business models, business strategy, and innovation. Long Range Planning, 43(2-3), 172-194.

Shamsaddini, R., Vesal, S. M., & Nawaser, K. (2015). A new model for inventory items classification through the integration of ABC–Fuzzy and fuzzy analytic hierarchy process. International Journal of Industrial and Systems Engineering, 19(2): 239-261.

Sepahvand, R., Nawaser, K., Azadi, M. H., Vafaei-Zadeh, A., Hanifah, H., & Khodashahri, R.B. (2023). In search of sustainable electronic human resource management in public organizations. Int. J. Information and Decision Sciences, 15(2), 117–147.

Vesal, S.M., Nazari, M., Hosseinzadeh, M., Shamsaddini, R., Nawaser, K. (2013). The relationship between labor market efficiency and business sophistication in global competitiveness, International Journal of Business and Management, 13 (8): 83-92.
15